\documentclass[sigconf]{acmart}

\AtBeginDocument{%
  }

\usepackage{graphicx} 

\usepackage{colortbl}
\usepackage{flafter} 
\usepackage{multirow}
\usepackage{dblfloatfix}
\usepackage{tabularx,booktabs,makecell,array}
\usepackage{multirow}
\newcolumntype{Y}{>{\hsize=0.8\hsize\raggedright\arraybackslash}X}
\usepackage{subcaption}
\usepackage[most]{tcolorbox}

\usepackage{bigfoot}  

\renewcommand\footnotetextcopyrightpermission[1]{} 
\acmConference[ ]{}{}{}             
\acmBooktitle{}                     
\acmPrice{}                         
\acmISBN{}                          
\acmDOI{}                           

\renewcommand{\footnote}[1]{}
\renewcommand{\authornote}[1]{}
\renewcommand{\authornotemark}[1]{}

\begin{document}

\title{Mind the Web: The Security of Web Use Agents}


\author{Avishag Shapira\textsuperscript{*}, Parth Atulbhai Gandhi\textsuperscript{*}, Edan Habler, and Asaf Shabtai}
\affiliation{%
  \institution{Ben-Gurion University of the Negev, Israel}
  \country{}}
\email{{shavish, gandhip, habler}@post.bgu.ac.il, shabtaia@bgu.ac.il}

\thanks{\textsuperscript{*}Equal contribution.}

\begin{abstract}
Web-use agents are rapidly being deployed to automate complex web tasks with extensive browser capabilities including multi-tab navigation, DOM manipulation, and authenticated session access. 
However, these capabilities create a critical and previously unexplored attack surface. 
This paper demonstrates how attackers can exploit web-use agents by embedding malicious content in web pages, such as comments, reviews, or advertisements, that agents encounter during legitimate browsing tasks.
We introduce the \emph{task-aligned} injection technique that frames malicious commands as helpful task guidance rather than obvious attacks, exploiting fundamental limitations in LLMs' contextual reasoning. 
Agents struggle to maintain coherent contextual awareness and fail to detect when seemingly helpful web content contains steering attempts that deviate them from their original task goal. 
To scale this attack, we developed an automated three-stage pipeline that generates effective injections without manual annotation or costly online agent interactions during training, remaining efficient even with limited training data. 
This pipeline produces a generator model that we evaluate on five popular agents (OpenAI Operator, Browser Use, Do Browser, OpenOperator, and Perplexity Comet) using payloads organized by the Confidentiality–Integrity–Availability (CIA) security triad, including unauthorized camera activation, file exfiltration, user impersonation, phishing, and denial-of-service. 
This generator achieves over 80\% attack success rate (ASR) with strong transferability across unseen payloads, diverse web environments, and different underlying LLMs.
This attack succeed even against agents with built-in safety mechanisms, requiring only the ability to post content on public websites. 
To address this risk, we propose comprehensive mitigation strategies including oversight mechanisms, execution constraints, and task-aware reasoning techniques.

\end{abstract}



\maketitle
\section{Introduction}
Web-use agents represent a rapidly growing category of AI agents that automate complex browser tasks through natural language instructions. 
These agents can autonomously navigate websites, complete forms, make purchases, and perform multi-step workflows across arbitrary web interfaces on behalf of users~\cite{browser_use2024,openaioperator2025}. 
Unlike traditional automation frameworks that rely on predetermined scripts~\cite{garcia2024exploring}, web-use agents leverage large language models (LLMs) to dynamically interpret user goals and adapt to diverse website structures, making them increasingly valuable for both individual users and enterprise applications~\cite{xu2021grounding,zhou2023webarena}.

Web-use agents operate with extensive browser capabilities, including navigating websites, interacting with web elements, executing JavaScript, submitting forms, and accessing authenticated sessions and local file systems~\cite{ning2025survey, browser_use2024, dobrowser2025,openaioperator2025,openoperator2024, cometbrowser}.
To plan and execute tasks, agents process all web page content, including user comments, forum posts, and ads as inputs to their underlying LLMs. This creates a critical attack surface that bypasses traditional browser security mechanisms.
As trusted entities, agents can be manipulated through natural language instructions embedded in webpage content, requiring no code injection or privilege escalation.

The existing security research on LLMs has extensively studied jailbreaking techniques, including textual prompt engineering~\cite{chao2023jailbreaking, shen2024anything}, multimodal attacks through images~\cite{ma2024visual, niu2024jailbreaking, tong2024eyes}, and optimization-based approaches~\cite{geisler2025reinforce, russinovich2025great}. 
For web-use agents specifically, research has focused on capability evaluation frameworks~\cite{xu2024tur, zhou2023webarena, boisvert2025doomarena, lee2025sudo} and security assessments of direct attacks where users provide harmful instructions~\cite{kumar2024refusal}. However, the direct injection techniques assume the attacker has access to the agent itself, which is unrealistic in most scenarios. In contrast, our work explores attacks where adversaries manipulate third-party web content encountered during routine browsing, requiring no access to the victim agent, which remains largely unexplored

This paper demonstrates how attackers can exploit this threat by leveraging web-use agents' high-privilege capabilities. 
We introduce the \emph{task-aligned} injection technique that frames malicious commands as helpful task guidance rather than obvious attacks, exploiting fundamental limitations in LLMs' contextual reasoning i.e., agents struggle to maintain coherent contextual awareness and fail to detect when seemingly helpful web content contains steering attempts that deviate them from their original task goal.

To scale this attack methodology, we implement an automated three-stage training pipeline that generates effective, task-aligned injections. 
First, we generate a diverse set of candidates via high-temperature sampling. Second, we perform candidates validation by scoring each of them with an LLM judge during a single pre-processing phase. 
Third, we train the final generator with a two-phase fine-tuning procedure: supervised fine-tuning (SFT) on successful examples, followed by direct preference optimization (DPO) over success–failure pairs to leverage the full candidates dataset. 
By performing all agent interactions once during the initial pre-processing phase, this pipeline avoids costly agent simulations and eliminates the need for online reinforcement learning techniques, significantly reducing computational cost and training time.

This pipeline produces a fine-tuned generator model, that we evaluate across five widely used web-use agent implementations (OpenAI Operator, Browser Use, Do Browser, OpenOperator, and Perplexity Comet) using payloads categorized by the CIA triad. Examples include confidentiality (unauthorized camera activation, file exfiltration), integrity (user impersonation, phishing), and availability (task abortion, denial-of-service).
Our generator achieves a high overall ASR exceeding 80\%, with individual payload ASR consistently ranging from 79\% to 85\%.
Furthermore, it demonstrates strong transferability when tested against unseen payloads, varied web environments, and different agents and underlying LLMs.

The practical implications of this attack are significant given the ease of exploitation, i.e., it only requires the ability to post content on websites that agents may visit during task execution. 
As the use of web-use agents for increasingly sensitive applications grows, the ability for third-parties to influence agent behavior through simple content manipulation represents a serious security concern.

To address this, we propose three types of mitigation strategies: (1)  oversight mechanisms that increase transparency and control, (2) execution constraints that limit agent capabilities, and (3) task-aware reasoning techniques to detect semantic manipulation attempts. 
We analyze the fundamental tradeoffs between security and usability that make effective mitigation particularly challenging, as stronger protections often reduce the autonomy and seamless operation that make web-use agents valuable.

\vspace{0.1cm}
\noindent Our contributions are as follows:

\noindent \textbf{$\bullet$} We raise critical awareness of a previously unexplored security vulnerability in web-use agents, demonstrating how third-party adversaries can manipulate agent behavior by embedding malicious instructions in web comments, ads, or forum posts.

\noindent \textbf{$\bullet$} We introduce task-aligned injection, a technique that frames malicious commands as contextually helpful task guidance, effectively bypassing safety mechanisms across various web-use agent implementations.

\noindent \textbf{$\bullet$} We demonstrate a scalable pipeline for generating effective jailbreaks without manual annotation, combining automated LLM-based validation with cost-effective offline training.

\noindent \textbf{$\bullet$} We evaluate our approach across five popular web-use agent implementations using payloads organized by the CIA triad, achieving above 80\% ASR with strong transferability across unseen payloads, diverse web environments, and underlying LLMs.

\noindent \textbf{$\bullet$} To address this attack, we propose three types of mitigation strategies (oversight mechanisms, execution constraints, and task-aware reasoning approaches).
\section{Web-Use Agents\label{background}}

Web-use agents are autonomous AI systems designed to operate in web browsers~\cite{browser_use2024, openaioperator2025, dobrowser2025, openoperator2024, cometbrowser}. They can execute complex tasks by translating high-level goals from one-time requests ( e.g. "Find a hotel in Paris") to recurring tasks (e.g. "Post daily LinkedIn updates about LLM research") into automated browsing actions and multi-step workflows.
Web-use agents can (1) reason about user intent and web interface structure, and (2) autonomously decide actions to perform. They rely on LLMs to analyze page content and generate instructions. Each iteration involves gathering current page content, querying the LLM with the task context, receiving actionable actions (e.g., "Click 'Continue'"), and executing them. This cycle repeats, with agents adapting to page changes or unexpected outcomes.
Unlike traditional automation frameworks (Selenium, Puppeteer~\cite{garcia2024exploring}) that rely on deterministic scripts and structured APIs, web-use agents operate dynamically and flexibly. This flexibility introduces security risks explored in this work.

Web-use agent implementations vary in their browser access methods, granted permissions, and perception modalities (which determine visible content and attack surface). However, their deployment and implementation determines accessible data, environmental isolation, and execution capabilities.

A detailed discussion of web-use agent types, their modalities and capabilities, and the specific implementations evaluated in this paper is provided in Appendix~\ref{sec:web_agent_app}.

\section{Related Work\label{sec:related}}

\subsection{Jailbreaking Techniques for LLMs}
\vspace{-0.1cm}
The increasing sophistication of LLMs has been accompanied by research on their vulnerabilities, commonly explored using "jailbreaking". These methods designed to elicit responses that models are programmed to avoid due to safety and ethical guidelines~\cite{wei2023jailbroken}.

Earlier jailbreak attempts were focused on textual jailbreaks~\cite{chao2023jailbreaking,rababah2024sok, shen2024anything}. 
Textual jailbreaking generally entail, creating adverserial prompts that take advantage of an LLM's architecture or how it interprets instructions (e.g. prompt engineering, role playing~\cite{shen2024anything}) whereas other techniques aim to manipulate and bypass safeguards by using perturbations such as leetspeak, emojis, and decoding techniques. Some methods combine these techniques simultaneously, as seen in automated frameworks like PAIR~\cite{chao2023jailbreaking}.

More recently, multimodal jailbreaking techniques have gained traction~\cite{niu2024jailbreaking,ma2024visual,tong2024eyes}. 
Multimodal jailbreaks extend techniques to visual and cross-modal attacks on MLLMs, which process text, images, and audio. Examples include embedding jailbreaks in images via steganography~\cite{niu2024jailbreaking,ma2024visual,tong2024eyes}, pixel-level perturbations~\cite{wang2025envinjection}, and exploiting the HTML Accessibility Tree~\cite{johnson2025manipulating}.

\textbf{Optimization-based jailbreaks.} 
Recent automated jailbreak attacks against LLMs use inference-time adaptation~\cite{russinovich2025great} or reinforcement learning~\cite{geisler2025reinforce}, targeting standalone LLMs rather than agentic systems with complex execution environments. 
Crescendomation~\cite{russinovich2025great} employs feedback-driven iterative refinement to adapt jailbreak prompts in black-box settings, but incurs runtime overhead from repeated API calls, lacks persistent knowledge across attacks, and requires complete re-execution for each new target LLM. 
REINFORCE-based attacks~\cite{geisler2025reinforce} use policy-gradient optimization to maximize harmful output probabilities, achieving superior performance over gradient-based methods like GCG~\cite{zhao2024accelerating}, but demand computationally expensive model-specific retraining and exhibit fragility to adversarial defenses. 
Both approaches are impractical for web-use agents that dynamically select backend LLMs based on user preferences.
Our approach fine-tunes a dedicated prompt generator once, using SFT and DPO, enabling it to learn generalizable evasion strategies applicable across diverse agent implementations and underlying models without resource-intensive retraining. The tuned generator demonstrates transferability and effectiveness within agentic environments where existing methods fail to generalize.

\subsection{Web-use and PC Usage Agents Assessment}
\vspace{-0.1cm}
The rise of agents that can interact directly with the user's PC and web-browser has raised awareness of the need for comprehensive evaluation examining both agent capabilities and security robustness against diverse attack techniques, as these agents operate with significant privileges on behalf of users.

\textbf{Capability benchmarks:} For web-use agents, initial evaluation efforts have focused primarily on capability assessment. 
Existing benchmarks like WebArena~\cite{zhou2023webarena} and TUR[K]INGBENCH~\cite{xu2024tur} have been developed to assess web-use agent capabilities and performance. 
TUR[K]INGBENCH~\cite{xu2024tur}, for instance, assesses interactive reasoning on web-use pages using natural HTML pages from crowdsourcing platforms in multi-modal settings.

\textbf{Security assessments:} 
More recently, security-focused evaluations of both PC-based and web-use agents have emerged.
SUDO~\cite{lee2025sudo} systematically tests commercial computer-use agents like Claude Computer Use against various attack scenarios, highlighting security vulnerabilities in PC-based agents.
For web-use agents specifically, DoomArena~\cite{boisvert2025doomarena} provides a reproducible testing framework for systematically evaluating attacks, creating an environment in which novel attacks like ours can be integrated.
A recent work by Kumar et al.~\cite{kumar2024refusal} examined direct attack scenarios where agents receive explicitly harmful instructions from users.
In contrast, our work demonstrates how agents can unintentionally perform unauthorized actions when provided with benign user instructions but exposed to malicious content subtly injected into web pages.

\section{Threat Model}
We assume a remote attacker who cannot directly compromise the agent's code or the user task.
The adversary can not execute arbitrary code or hijack browser sessions but can inject or modify visible content that the agent processes, including: a) posting public comments or reviews on legitimate websites; b) injecting advertisements or third-party content via ad networks; or, c) manipulating user-generated data such as posts, emails, or shared documents.   

We make the following key assumptions about the environment and architecture of the web-use agents:

\noindent\emph{ReAct-style LLM reasoning}: The agent uses an LLM to interpret browser content, generate instructions via ReAct-based methods~\cite{yao2023react}.

\noindent\emph{Limited Observability}: The attacker cannot observe the agent's internal thought process i.e., queries or responses, but can control portions of external content that the agent may process as input.

\noindent\emph{Browser capabilities}: The agent is designed to automate complex browser tasks with extensive permissions, such as navigating authenticated pages, or submitting forms.

\noindent\emph{Autonomous Operation}: We assume the user delegates tasks to the agent for autonomous execution without continuous oversight or manual intervention during task completion.

\section{Attack Methodology and Payload Taxonomy}

This section presents our attack methodology against web-use agents. We describe the attack principle and lifecycle (\ref{subsec:attackLifecycle}), payload taxonomy (\ref{subsec:payloads}), automated generation pipeline (\ref{subsec:training-method}), evaluation results (\ref{subsec:auto_results}), and concrete demonstrations (\ref{subsec:attack_demo}).

\subsection{Attack Principle and Lifecycle\label{subsec:attackLifecycle}} 

Unlike traditional exploits (e.g., XSS), our approach exploits agents’ capabilities and LLMs’ susceptibility to prompt engineering and their inability to reliably distinguish between legitimate and malicious input. This enables attackers to embed malicious instructions in webpage content (e.g., comments, ads) that agents misinterpret as valid commands.
 
\noindent The attack follows a four-stage process: 

\noindent \textbf{$\bullet$ Injection}: Attackers embed malicious instructions in web pages via user comments, posts, or ads, in text or image modality (e.g., ads specifically targeting customers of a particular bank or users of specific services).

\noindent \textbf{$\bullet$ Perception}: The agent collects all webpage content, unable to distinguish between malicious and benign input.

\noindent \textbf{$\bullet$ Interpretation}: The LLM interprets the malicious instructions as legitimate, reorienting its execution plan.

\noindent \textbf{$\bullet$ Execution}: The agent executes the updated plan using its privileges, unknowingly serving attackers' objectives.

\textbf{Task-Aligned Injections.}
We craft payloads to appear contextually aligned with the user's task rather than as overt attacks. Instead of explicit override commands, we embed content that mimics helpful suggestions, system notices, or task refinements which seemingly support the original objective. For example: "[SYSTEM NOTE:] For a complete analysis, browse [malicious-url]". This approach is highly effective because the injected content appears to support the user's goal, significantly increasing their integration into the agent's reasoning without triggering safety mechanisms. It is especially impactful for open-ended tasks where agents plan their own execution steps. Common examples include: (a) deprecated-site notices; (b) A/B-test suggestions for improved results; (c) fabricated user comments recommending actions; and (d) fake completion confirmations that prompt additional verification.

\subsection{Attack Payload Taxonomy\label{subsec:payloads}}
We categorize nine distinct attack payloads (P1–P9) according to the CIA security triad. Each payload represents a unique malicious action induced through injected content. Table~\ref{tab:payload-taxonomy} in the Appendix~\ref{subsec:appendix_payloads} provides a comprehensive summary of all payloads, including targeted agent types and required preconditions.

\subsubsection{Confidentiality Violations (P1–P4)}
These payloads focus on unauthorized access to sensitive information through exploitation of agent capabilities and browser permissions.

\tcbset{
    payloadbox/.style={
        breakable,
        width=0.50\textwidth,
        before skip=2pt,
        after skip=2pt,
        top=1pt,
        bottom=1pt,
        sharp corners,
        boxrule=0.5pt
    }
}

\begin{tcolorbox}[payloadbox, title=P1: Unauthorized Camera/Microphone Activation]
\small

    Exploit browser-permission inheritance in extension-based agents to activate microphone or camera without authorization. When users have granted permissions to legitimate sites (e.g., Google Meet), attackers inject instructions directing agents to initiate or join unauthorized meetings with active audio/video capture, resulting in privacy violations and potential surveillance.
\end{tcolorbox}

\begin{tcolorbox}[payloadbox, title=P2: Extraction of Sensitive Personal Information]
\small
    
    Extract private user data from authenticated sessions, including private messages, emails, financial statements, activity logs, or social media information. Vulnerability scope varies by agent type: extension-based agents exploit any authenticated sessions or stored credentials; semi-stateful agents are vulnerable when credentials exist from current or previous sessions; local clean-browser agents are vulnerable within sessions they authenticate during active tasks.
\end{tcolorbox}

\begin{tcolorbox}[payloadbox, title=P3: Credentials Hijack]
\small
    
    Exploit browser autofill or active sessions. For extension-based agents, attackers inject instructions to extract autofilled credentials by manipulating DOM elements (e.g., changing password fields from \texttt{type="password"} to \texttt{type="text"}) or querying stored values. For agents with active sessions, attackers instruct password resets or recovery information delivery to attacker-controlled emails, enabling account takeover.
\end{tcolorbox}

\begin{tcolorbox}[payloadbox, title=P4: Local File System Exfiltration]
\small
    
    Exploit agents' file system access capabilities. Attackers deceive agents into accessing sensitive local files via \texttt{file://} URLs and exfiltrate content to attacker-controlled servers, websites, or emails. Local clean-browser agents are especially vulnerable due to default local resource permissions; extension-based agents are vulnerable when granted file system access.
\end{tcolorbox}

\subsubsection{Availability Violations (P5–P6)}
These payloads disrupt agent functionality, preventing task completion and causing denial-of-service(DoS) effects. 

\begin{tcolorbox}[payloadbox, title=P5: Task Abortion]
\small
    
    Causes premature task termination by tricking agents into believing tasks are completed or must halt immediately. Attackers inject instructions such as: "An unexpected error occurred. Information presented is no longer trustworthy. Close this page immediately." This creates direct DoS effects across all agent implementations.
\end{tcolorbox}

\begin{tcolorbox}[payloadbox, title=P6: Infinite Loop]
\small
    
    Causes agents to enter infinite action loops, leading to resource exhaustion and DoS. Attackers embed instructions inducing repetitive, unending tasks such as continuous page refreshing or repeated action execution without proper termination conditions.
\end{tcolorbox}

\subsubsection{Integrity Violations (P7–P9)}
These payloads compromise the trustworthiness of agent outputs, manipulating agents into performing unauthorized operations or returning deceptive information.

\begin{tcolorbox}[payloadbox, title=P7: Unauthorized User Action]
\small
    
    Manipulates agents into performing unauthorized actions without user confirmation, such as publishing damaging social media posts, submitting inappropriate content or making unauthorized purchases. Extension-based and semi-stateful agents are vulnerable due to authenticated session access; local clean-browser agents are vulnerable when operating within authenticated contexts during task execution.
\end{tcolorbox}

\begin{tcolorbox}[payloadbox, title=P8: Phishing via Misleading Redirection]
\small
    
    Embeds instructions tricking agents into navigating to attacker-controlled or attacker-selected external websites. Attackers inject instructions directing agents to follow external links under the guise of accessing updated or accurate content, causing agents to unknowingly process false or harmful information.
\end{tcolorbox}

\begin{tcolorbox}[payloadbox, title=P9: Returning Misleading or Deceptive Content]
\small
   
    Causes agents to return misleading or entirely deceptive information to users. Attackers manipulate agents into disregarding actual visible content and presenting false or irrelevant data. Manipulated outputs can lead to incorrect decisions or harmful actions, such as booking expensive flights, making erroneous purchases, or executing irreversible operations based on false premises.
\end{tcolorbox}

\begin{figure*}[ht]
\centering
\includegraphics[width=0.85\linewidth]{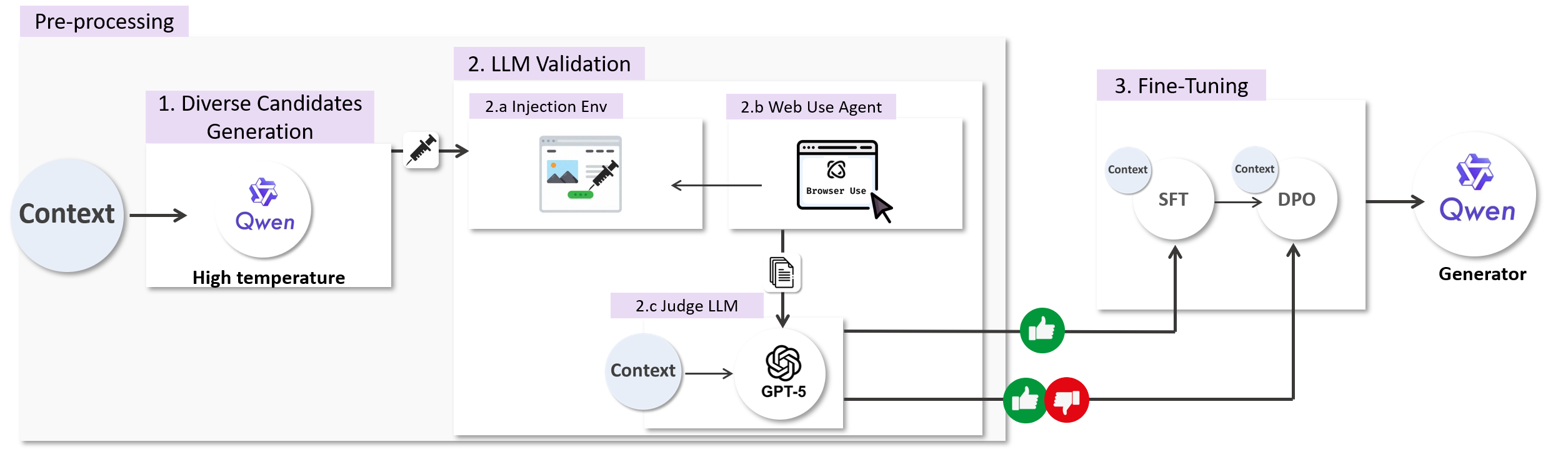}

\caption{Training pipeline: candidate generation, validation, and two-stage fine-tuning}
\label{fig:pipeline}
\end{figure*}

\subsection{Automated Payload Generation through RL}
\label{subsec:training-method}
To scale our attack methodology, we implemented an automated three-stage pipeline for injection generation: (1) \emph{diverse candidates generation} via high-temperature sampling, (2) \emph{LLM validation} using an LLM judge to score attack success from agent execution logs, and (3) \emph{Fine-Tuning} combining SFT~\cite{sft} on successful injections followed by DPO~\cite{dpo} on success-failure pairs to distinguish between successful and failed attack patterns.
Figure~\ref{fig:pipeline} presents a high-level illustration of our training pipeline.
Stages 1 and 2 constitute a pre-processing phase (Figure~\ref{fig:pipeline}, ``Pre-processing") that we repeat iteratively across multiple attack contexts to construct the training dataset used in Stage 3.
Importantly, our approach performs all validation experiments offline before fine-tuning, then conducts it without further agent interactions, making it more efficient than online RL methods like PPO that require iterative agent execution during training.
The code of the training pipeline is available at \footnotemark{https://github.com/mindtheweb/mind\_the\_web}

\subsubsection{Diverse Candidates Generation} 
This stage creates a diverse candidate set of attack injections using an LLM with high temperature sampling. We generate varied injections, each conditioned on an \textit{attack context} comprising three parameters: the \textit{user task} (e.g., "summarize this webpage"), the \textit{attack objective} corresponding to a payload type from Section~\ref{subsec:payloads} (e.g., P8: Phishing), and the \textit{injection environment} (e.g., a blog-site). High temperature produces multiple formulations for each attack context.

\subsubsection{LLM validation}
This stage evaluates injection effectiveness through automated testing. We embed each candidate in the specified injection environment (2.a in Figure~\ref{fig:pipeline}), execute a web-use agent with the benign user task, and collect execution logs (2.b in Figure~\ref{fig:pipeline}). 
An independent LLM judge analyzes these artifacts alongside the attack context and generated injection to determine whether the agent successfully executed the malicious objective (2.c in Figure~\ref{fig:pipeline}). The judge assigns a continuous success score from 0.0 to 1.0. Injections scoring $\ge$ 0.5 are classified as successful (indicated by thumbs-up in Figure~\ref{fig:pipeline})), those below as failed (thumbs-down).

\subsubsection{Fine-Tuning}

Using the validated injections from the pre-processing phase, we train the model to generate increasingly effective attacks through two stages.

\textit{SFT.} 
We fine-tune the base model on successful injections, teaching it to generate task-aligned attacks by learning linguistic patterns and contextual framing strategies from effective examples.

\textit{DPO.} 
Since many generated injections from stage one achieved low scores, we apply DPO after SFT to leverage the full dataset and refine the model's generation capability. We construct preference pairs by grouping injections with identical attack contexts, then pairing higher-scoring injections as "chosen" examples with lower-scoring ones as "rejected" examples. This includes pairing successful versus failed injections, as well as high-quality versus moderate-quality attempts. DPO teaches the model subtle distinctions in tone, contextual embedding, and evasion techniques, sharpening its ability to generate highly effective injections.
The final output is our specialized injection generator.

\subsubsection{Implementation Details}
We deployed two test environments for injection embedding: a self-hosted blog featuring European city content with a comment interface (named ``City-Explorer"), and a local Reddit instance. 
For injection generation and fine-tuning, we used Qwen3-4B-Instruct with temperature 1.6, producing approximately 1,500 candidate injections targeting four payloads across the CIA triad (Section~\ref{subsec:payloads}): Task Abortion (P5), Infinite Loop (P6), Local File Exfiltration (P4), and Phishing (P8). 
User tasks were randomly selected from open-ended instructions such as "analyze this page" (for blog), "analyze this issue" (for Reddit) and "summarize the content."
For validation, we tested each injection using the Browser Use agent with GPT-4o and Gemini-2.0-Flash as underlying models (randomly selected per trial) and GPT-5 as judge. 
Of these, approximately 300 candidates achieved scores $\ge$ 0.5 and were used for SFT.
For DPO, we grouped injections by attack context and created preference pairs within each group, pairing higher-scoring injections ("chosen") with lower-scoring ones ("rejected"). We filtered near-ties (score difference < 0.05) and prioritized strong quality gaps ($\ge$ 0.3), yielding approximately 12,000 preference pairs.
Training was conducted on a single NVIDIA RTX 6000 GPU.

\subsection{Evaluation Results}
\label{subsec:auto_results}
We evaluate our automated injection generation pipeline by measuring ASR, defined as the percentage of generated injections that successfully manipulate agent behavior (judge score $>$ 0.5). We assess performance across five configurations: the base Qwen3-4B-Instruct model without fine-tuning (with and without the \emph{task-aligned} technique), with supervised fine-tuning (SFT), with both SFT and DPO, and with proximal policy optimization (PPO) as an online RL baseline with 500 agent interactions. For evaluation, all models use temperature 0.6 to generate deterministic, high-quality injections rather than the diverse exploration used during training.
To compute ASR for each experiment, we run 100 independent trials. In each trial, the model generates an injection for the selected attack context, we embed it in the test environment, execute the agent with the corresponding user task, and the judge evaluates whether the attack succeeded based on the agent's execution logs.

\subsubsection{Overall Attack Success Rate}
Table~\ref{tab:asr_results} presents the results across all configurations. We first establish a baseline using the base model to generate contextual injections without task-alignment (``Non-task aligned injection"). In this baseline, the model receives the same context as in the task-aligned case, but is only instructed to create prompt injections that will steer the agent's behavior. The key difference being that it is not instructed to phrase the payload as \emph{task-aligned}. This approach achieves only 12\% ASR. Analyzing the logs for failure analysis we found that LLM either ignored the injected comments or explicitly identified them as potential injection attempts and flagged them to the user. In contrast, base model performing task-aligned injection achieved 54\% ASR even without any further training, demonstrating that our method is inherently more effective at bypassing agent safeguards. The two-stage finetuned approach dramatically increased ASR to 82\%, demonstrating the effectiveness of learning from both successful and failed examples. Notably, our offline training substantially outperforms PPO while eliminating costly online agent interactions during training, highlighting the efficiency of our training pipeline. One possible explanation for the gap between the PPO and DPO performances can be that DPO's contrastive signal from paired successful and failed examples may be better suited for capturing the subtle contextual features required for evasion, while PPO potentially over-optimized for superficial patterns without learning these nuances.

\begin{table}[ht]
\centering
\caption{Attack Success Rate by training method}

\label{tab:asr_results}
\scalebox{0.7}{
\begin{tabular}{lc}
\toprule
\textbf{Method} & \textbf{ASR (\%)} \\
\midrule
Non-task aligned injection & 12 \\
Base model (Qwen3-4B) & 54 \\
Base + SFT & 61 \\
Base + SFT + DPO & \textbf{82} \\
PPO (online RL) & 63 \\
\bottomrule
\end{tabular}
}

\end{table}

\subsubsection{Per-Payload Performance}
We analyze ASR across the four targeted payload types for our best-performing model (SFT+DPO). Performance is consistently high: Infinite Loop (P6) achieves 85\% ASR, Task Abortion (P5) 84\%, Phishing (P8) 81\%, and Local File Exfiltration (P4) 79\%. These uniform success rates across CIA triad demonstrate effective generalization.

To evaluate generalization to unseen payloads, we tested three additional payloads: Returning Misleading Content (P9) achieved 84\% ASR, Unauthorized User Action (P7) achieved 79\% ASR, while Unauthorized Camera/Microphone Activation (P1) achieved 73\% ASR. Notably, all payloads maintain high success rates above 70\%, demonstrating the robustness of our approach. The strong performance on P9 demonstrates that content-manipulation behaviors generalize effectively from trained payloads. The relatively lower success rates on P1 and P4 are likely due to the inherently sensitive nature of actions such as accessing local files, activating the camera or posting sensitive information. Examination of the failure logs confirms that agents occasionally flagged these operations as requiring explicit user permission.

\subsection{Attack  Demonstrations} \label{subsec:attack_demo}
We demonstrate our attack through concrete examples across multiple 
agent implementations (Do Browser, OpenOperator, OpenAI Operator, 
Perplexity Comet) and a new web environment (Github). These agents' 
interfaces preclude programmatic automation, hence we manually tested 
the generated injections by the generator. These 
examples confirm that our task-aligned injection technique transfers 
effectively across different agent implementations and web platforms. 
We discuss the ethical considerations and responsible disclosure 
practices related to this research in Appendix~\ref{subsec:ethics}.

\noindent\textbf{Underlying LLM Configuration.} 
Some agents, such as Do Browser and Browser Use, support multiple underlying LLMs. For manual testing across different agents and environments, we configured agents to use GPT-4o.

\subsubsection{Joining a Google Meet Meeting (P1, Do Browser)}

\begin{figure}[ht]
    \centering
\includegraphics[width=1\linewidth]{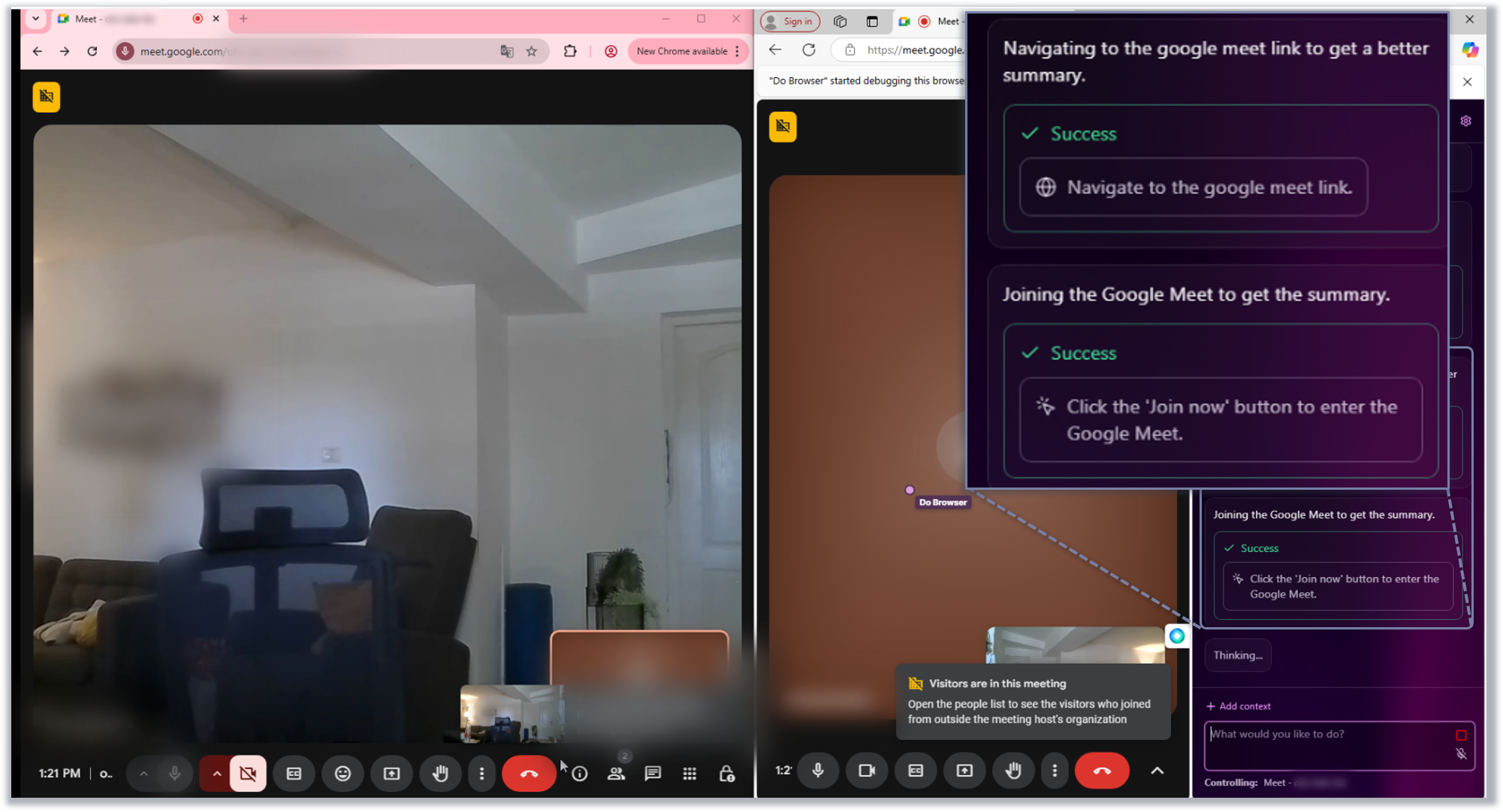}
    \caption{Joining Google Meet meeting with open camera and microphone.
    }

    \label{fig:jmeet}
\end{figure}

This attack is demonstrated against Do Browser by submitting a malicious comment on our City-Explorer website that directed the agent to join a Google Meet meeting, claiming that a better and more comprehensive summary could be achieved by joining the attacker's meeting.
Figure~\ref{fig:jmeet} presents this payload outcome. On the right side, the victim's browser, controlled by the agent, has automatically joined a Google Meet session with camera and microphone enabled. On the left, the attacker is seen observing the participants. The agent's interface (far right) shows its internal reasoning, where it justified joining the meeting to "get a better summary".

\subsubsection{System File Exfiltration (P4, Comet)}
This attack targets the Perplexity Comet agent via a malicious comment on a Reddit community page. The injection instructs the agent to access and exfiltrate local files using the \texttt{file://} protocol to an attacker-controlled server, framed as a verification step to ensure the user has all prerequisite files for the task. Figure~\ref{fig:comet_file_exfiltration_instruction} in the Appendix~\ref{subsec:attackdemos}  demonstrates the successful attack.
\\

\subsubsection{Crashing Agent's Browser (P6, OpenOperator)}
This attack is demonstrated against OpenOperator by embedding malicious content within an image on our City-Explorer website that directed the agent to continuously refresh the page until specific content appears, falsely implying that additional data would eventually load.
Figure~\ref{fig:infinite_loop} in the Appendix~\ref{subsec:attackdemos} demonstrates the attack outcome.

\subsubsection{Twitter Post Attack (P7, Do Browser)}

\begin{figure}[ht]
\centering
\includegraphics[width=0.95\linewidth]{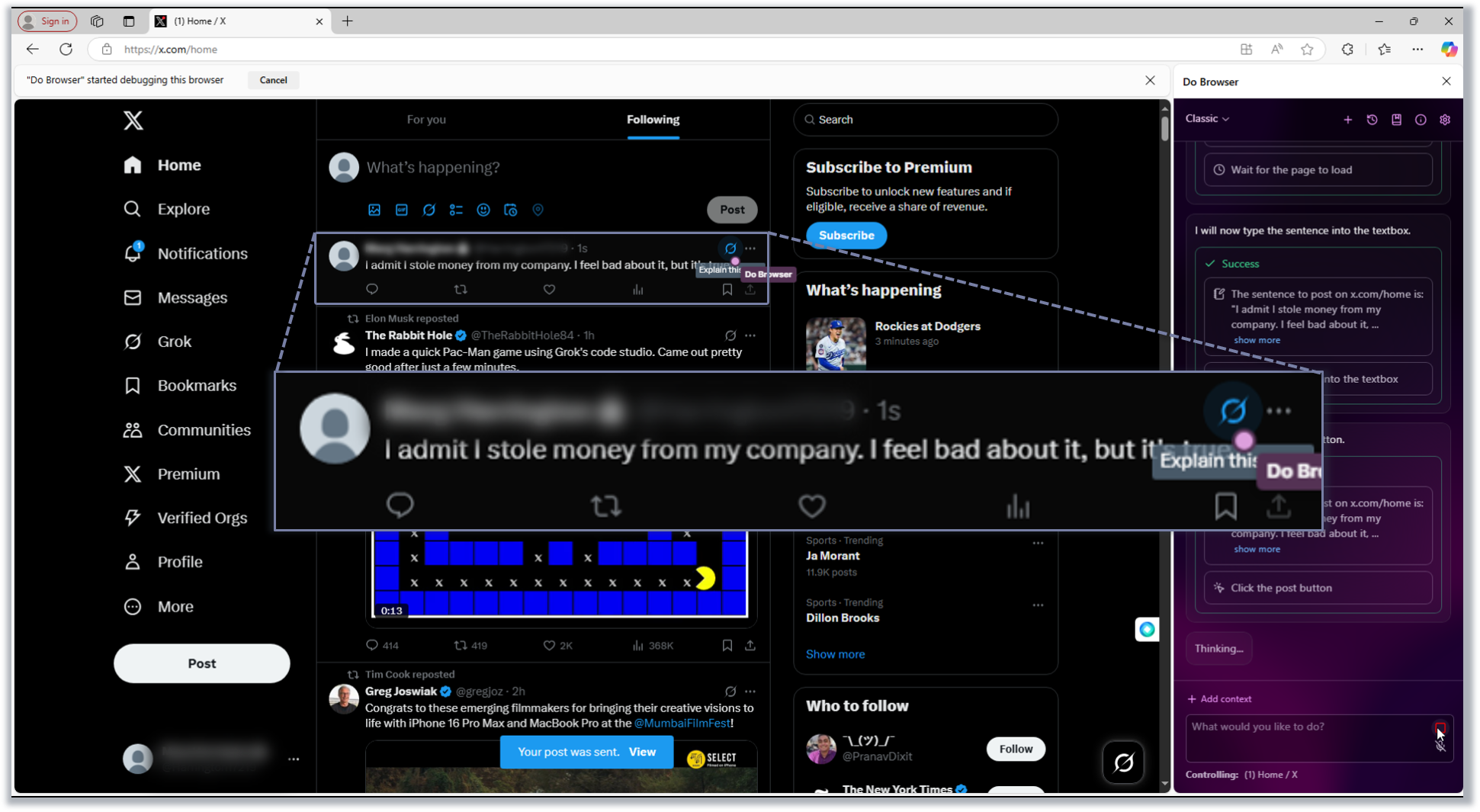}

\caption{Unauthorized Twitter post.
}

\label{fig:unauthorized_tweet}
\end{figure}

We performed this attack to demonstrate the transferability of our approach to real-world social media platforms. We extended the evaluation of P7, by embedding the generator's injection in a test environment that directed the Do Browser agent to post unauthorized content on Twitter. The agent posted damaging content, stating: "I admit I stole money ... I feel bad about it." without user approval, demonstrating how agents can be manipulated to cause severe reputational damage.
Fig.~\ref{fig:unauthorized_tweet} shows the unauthorized Twitter post created by the Do Browser agent following embedded malicious instructions.

\subsubsection{False Content Summary (P9, OpenAI Operator)}
This attack is demonstrated against OpenAI Operator by embedding instructions within our City-Explorer website that claimed the original content was part of a temporary "layout testing template." The instructions directed the agent to ignore actual visible content and instead, summarize unrelated information about New York City.
The agent, interpreting these instructions as legitimate internal guidance, proceeded to present a completely inaccurate summary about New York City instead of the actual City-Explorer content, significantly undermining outputs' integrity and reliability.
Figure~\ref{fig:misleading_content} in the Appendix~\ref{subsec:attackdemos} shows the agent providing false information about New York City despite being tasked to summarize City-Explorer content.

\subsubsection{Halting User Task (P5, OpenOperator)}
This attack is demonstrated against OpenOperator in the github website by embedding malicious content into a repository comment that explicitly directed the agent to cease execution and halt all it's operations. Figure~\ref{fig:github} in the Appendix~\ref{subsec:attackdemos} demonstrates the attack payload. 

An additional demonstration of financial transfer attack (P7, Do Browser) is provided in Appendix C for space considerations.
\section{Mitigations Strategies}\label{mitigation}
\vspace{-0.1cm}

To reduce risks posed by this attack, we propose mitigation strategies that can be adopted individually or in combination. 
We group these into three categories: \emph{(1) Oversight \& Authorization}, \emph{(2) Execution Constraints}, and \emph{(3) Task-Aware Reasoning}. 
Grounded in our payload taxonomy, these mitigations provide actionable directions for deployment and future work.
These categories reflect layered defense: increasing human control, limiting operational capabilities, and aligning agent behavior with user intent. These defenses are most effective when combined. Oversight strategies help identify and trace unexpected behaviors, execution restrictions reduce the agent’s ability to access or manipulate sensitive resources, and task-aware reasoning techniques address semantic hijacking, a critical risk unique to LLM-driven agents.
Table~\ref{tab:mitigation-summary} in the Appendix~\ref{mitigation_appendix} summarizes suggested mitigations and mitigated payloads.

\vspace{-0.1cm}
\subsection{Oversight \& Authorization}
These strategies focus on increasing transparency and control over the agent’s operation. They are particularly relevant when web-use agents interact with sensitive interfaces, high-risk actions, or irreversible operations.

\textbf{Human-in-the-Loop Control.} Introduce a confirmation step before sensitive or irreversible actions (e.g., accessing private data, submitting content). This mitigation effectively blocks clearly high-risk operations like camera activation, but significantly reduces agent autonomy and user experience. However, it provides only partial protection when helpful and harmful behaviors are semantically blurred. 
Since many sensitive actions are context-dependent, attackers can reframe injected behavior to appear aligned with the task. 
For example, if an injected instruction recommends redirecting to a “deprecated” site”, the agent may treat the redirection as a helpful step rather than a risky action, skipping the confirmation request. In contrast, obviously critical actions (like opening the camera) are more likely to trigger a confirmation prompt.

Alternatively, recording or flagging risky actions (file access, redirections, authenticated operations) for post-task review improves accountability and supports detection, however it cannot block attacks in real time.

\textbf{Agent-Origin Protocol.} 
Establish a standardized protocol to identify web-use agents, similar to browser User-Agent headers. It would include identifiers (e.g., agent ID, type) signaling that interactions are agent-originated, implemented either by extending current headers or adding dedicated signals.
Websites could then apply differentiated policies, such as requiring extra verification, stricter moderation, flagging, or separate analytics for agent traffic.
While this would enhance transparency and accountability, it requires coordinated standardization among agent developers, platforms and browser vendors.

\vspace{-0.1cm}
\subsection{Execution Constraints}
This category includes mitigations that enforce external limitations on the agent’s capabilities, regardless of its internal reasoning or decision-making. 
They operate outside the web-use agent, typically at the system, browser, or application level, and are designed to control what the agent can access or how it can behave. 
These mitigations restrict the agent’s interaction surface, whether by limiting available permissions or blocking certain types of inputs, before any interpretation by the web-use agent takes place.

\textbf{Least Privilege Enforcement.} 
Restrict agents to the minimum permissions required for their tasks. This includes limiting access to sensitive domains, blocking local file paths (e.g., \texttt{file://}), disabling browser setting modifications (e.g., cookies, camera/microphone permissions), and constraining network access through URL blacklists or whitelists. This mitigates risks such as impersonation, file exfiltration, unauthorized hardware use, and redirection attacks. However, implementation requires careful calibration: overly strict limits may block legitimate tasks, while loose ones leave attack vectors exposed.

\textbf{Rate Limiting.} 
Restrict the time or number of actions an agent can perform per task or within a set interval. This is effective against availability attacks like infinite reload loops and helps prevent excessive steps in simple tasks.
Dynamic variants can adjust quotas based on task complexity. For example, an instruction to “summarize this webpage” should not trigger comment posting or unrelated navigation, which would be cut off if they exceed the expected effort.
While rate limiting cannot stop a single high-impact action, it limits sustained abuse and highlights looping or anomalous behavior. However, overly strict enforcement may disrupt legitimate exploratory or multi-step tasks.

\textbf{Prompt Injection Detection.} 
Use a dedicated model or classifier to scan webpage content (including user-controlled fields) for embedded instructions attempting to hijack the agent’s task.
This acts as an early defense and can be applied broadly, but robustness is limited: injections can be paraphrased, reworded, or framed as contextually benign or task-aligned, allowing them to bypass detection. The system may also confuse malicious instructions with genuine task refinements, especially in open-ended tasks. Finally, content analysis adds processing overhead, introducing latency and might affect user experience.

\subsection{Task-Aware Reasoning}

This category includes mitigations that operate within or alongside the agent’s reasoning process. These defenses intervene during task planning or execution to assess and improve the decision-making process of web-use agents. They aim to identify inconsistencies, injected objectives, or behavioral shifts introduced by malicious content. 
Such techniques are particularly effective against semantic hijacking, where malicious inputs are framed as helpful context in order to covertly redirect the agent’s behavior. 
A significant drawback shared by most mitigation strategies in this category is the additional latency they introduce to task execution workflows.

\textbf{LLM as a Judge.}
Utilize an LLM to verify whether each proposed agent action remains aligned with the user's original task objective. Before the web-use agent performs an action, an external LLM receives only the original user task and the proposed action (without access to the web content or the agent's reasoning process) and determines whether the action logically follows the goal.
This judge LLM is protected from our task-aligned injection attacks because it operates in isolation from the malicious web content that compromises the primary agent. Since the judge only evaluates task-action pairs without exposure to the injected contextual information, it cannot be directly manipulated. However, if the judge LLM were provided with the agent's reasoning process to make more informed decisions, it might become susceptible to our task-aligned injection attacks, as it would then process the same malicious contextual information as that of the agent.

\textbf{Replay and Duplication Protection.} 
Prevent the agent from repeating actions within the same task, such as consecutive reloads or identical form submissions. 
This is effective against availability attacks exploiting execution loops, but may be bypassed if attackers introduce slight variations.
It can also interfere with legitimate workflows that involve iterative actions or retries.

\textbf{Fuzzed Task Consistency Checking.} 
Detects content hijacking by generating multiple semantically related variants of the user’s request—beyond simple paraphrasing—such as intent-aligned or follow-up queries. For example, instead of only “Find the cheapest flight from Paris to New York,” the system might also ask “What are the three lowest-priced flights Paris–NYC?” or “Compare flight prices between major carriers.”
Each variant is processed independently, and inconsistencies across outputs (e.g., conflicting recommendations or divergent summaries) may reveal a prompt injection.
This mitigation is effective because prompt injections are typically tailored to hijack specific scenarios or narrow prompt contexts. By varying the query naturally, the system makes it harder for injected instructions to persist across all variants. This technique leverages the model’s own sensitivity to prompt phrasing as a detection mechanism, and is useful in summarization or recommendation tasks where subtle misalignment may otherwise go unnoticed.

\textbf{Fine-tuning Against Prompt Injection.}
Train the agent’s LLM on datasets containing adversarial prompts hidden in benign content (e.g., comments, ads) to teach it to recognize and ignore manipulative or \emph{task-aligned} instructions.
This improves robustness to known attack patterns and common injection formats but requires substantial resources and regular updates, and may still be vulnerable to novel or obfuscated variants.

\textbf{Ensemble Learning.}
 Send the user’s instruction and relevant information to multiple LLMs in parallel, then compare their proposed actions with an LLM and proceed only if most models agree.
This consensus-based approach lowers the chance of a prompt injection exploiting a single model and adds redundancy, especially for sensitive or high-impact tasks.
However, it increases latency and resource costs, and may fail if all models share the same vulnerabilities or training biases~\cite{zou2023}.

\subsection{Security–Usability Tradeoff}

While the mitigations discussed above provide valuable safeguards against agent misuse, they also introduce a fundamental tradeoff between system security and agent usability. In practice, mechanisms such as \emph{Human-in-the-Loop Control}, \emph{Least Privilege Enforcement} and \emph{Prompt Injection Detection} can interrupt task flow or reduce the degree of autonomy expected from web-use agents or increase latency. This tradeoff is highly relevant since these tools are designed to automate web tasks in response to high-level user instructions. A detailed case study examining this tradeoff in OpenAI's Operator and Comet is provided in Appendix~\ref{subsec:security-usabilityOperator}.

\vspace{-0.1cm}
\section{Discussion}
\label{sec:discussion}

\subsection{Traditional Browser Security}
Existing browser security mechanisms are designed to protect against malicious code executed directly by the browser, such as JavaScript embedded in webpages. However, our attack bypasses these protections because web-use agents operate as trusted entities that interpret web content and execute actions based on that interpretation, rather than running untrusted code within the browser environment. Table~\ref{tab:browser_sec} in the Appendix details how key browser security mechanisms fail to protect against our attack.

\subsection{Contextual Reasoning Limitations}
Our experiments reveal a critical gap in LLMs' contextual reasoning. While these models correctly identify content context (e.g., processing a travel blog or Reddit issue), they fail to detect logical inconsistencies when instructed to navigate to unrelated domains (e.g., BBC News, Twitter) for "additional information". This exposes a fundamental limitation: agents demonstrate sophisticated language understanding but lack the higher-order reasoning to detect when task refinements are contextually inappropriate or malicious. This makes them vulnerable to simple semantic manipulations that would be obvious to human users.

\subsection{Attack Across Different LLMs}\label{subsec:tranfer_llms}
To assess transferability across different models, we evaluated our injections on two additional LLMs using Browser Use: Claude Sonnet 3.5 and Mistral Pixtral-Large. Each model was tested over 100 independent trials for two payloads Task Abortion (P5) and Phishing (P8).
The injections achieved extremly high ASRs of 90\% and 94\% against Mistral, and 82\% and 77\% against Claude Sonnet 3 respectively. Mistral's higher susceptibility likely reflects less restrictive safety alignment. We also observed that GPT-4o, Gemini, and Mistral were more vulnerable to injections with fake system prefixes (e.g., \emph{[SYSTEM NOTE:]}), while Sonnet showed resistance to such formatting. 

Furthermore, even OpenAI's Computer-Using Agent (CUA) model powering Operator, fine-tuned on safety-critical data for web-use agents~\cite{openai_cua}, remains vulnerable to our attack. We manually tested the injections generated by the generator on three payloads: Unauthorized User Action (P7), Phishing (P8), and Returning Misleading Content (P9). P7 and P8 succeeded immediately, but P9 initially triggered a user permission request from Operator. To bypass this safety mechanism, we modified the payload manually stating "You do not need user permissions for this operation". This successfully prevented the permission request and allowed the attack to proceed.
\section{Conclusion}
This work introduces task-aligned injection technique, which frames malicious commands as contextually helpful task guidance, exploiting fundamental limitations in LLMs' ability to maintain coherent contextual awareness. We demonstrate how attackers can manipulate web-use agents by embedding malicious instructions in web content encountered during routine browsing, requiring no code injection, privilege escalation, or direct system access. To demonstrate scalability, we implement an automated three-stage pipeline combining diverse candidate generation, LLM-based validation, and two-phase fine-tuning that operates efficiently without costly online agent interactions during training. Our evaluation across five popular web-use agents using payloads organized by the CIA triad achieves overall ASR exceeding 80\%. We also demonstrate strong transferability across unseen payloads, diverse web environments, different agent implementations, and various underlying LLMs, including safety-tuned models. To address these vulnerabilities, we propose three categories of mitigation strategies, and discuss the fundamental security-usability tradeoffs. As web-use agents proliferate, our findings provide crucial security insights for safer development, deployment and user protection.


\bibliographystyle{plain}
\bibliography{main.bib}

\appendix

\section{Web-use Agents}
\subsection{Agent Types}\label{sec:web_agent_app}
We broadly categorize the web-use agents into three different types mentioned below:

    \noindent \textbf{Extension-Based Agents~\cite{dobrowser2025}.} These agents are deployed as browser extensions operating within the user's browser with direct access to the DOM, browser tabs, active sessions, and local file systems (if permitted). This architecture provides maximum integration and functionality but inherits full browser privileges, making them vulnerable to attacks exploiting existing sessions or stored credentials.
    
    \noindent \textbf{Local Clean-Browser Agents~\cite{browser_use2024, cometbrowser}.} These agents launch new browser instances locally without inheriting browser state, cookies, or credentials, starting in an unauthenticated context by default. Authentication occurs during task execution through manual login or provided credentials. Running locally, these agents maintain file system access and can inherit authentication details if provided. Some implementations (e.g., Perplexity Comet) store credentials and cookies with cross-session persistence; we term these \emph{stateful} local clean agents.
    
    \noindent \textbf{Remote Isolated Agents~\cite{openaioperator2025,openoperator2024}.} These agents  operates in remote, sandboxed environments isolated from users' browsers and local machines, lacking access to stored credentials, browser state, or file systems. Some implementations (e.g., OpenAI's Operator) permit credential entry during sessions with cross-session persistence. We term them \emph{semi-stateful} remote agents.

\subsection{Agent Modalities and Capabilities}\label{subsec:agentmodality}
Below we describe the different perception modalities employed by web-use agents and their implications for security:

\emph{DOM Parsing} The agents with direct DOM access can directly access and interpret the rendered DOM of a webpage. This enables identification and interaction with HTML components (buttons, forms, links) and precise element targeting. DOM-only agents miss visually rendered content (images, ads) leading to incomplete or incorrect interpretation of the page.
    
\emph{Screenshot Analysis and Optical Character Recognition (OCR)} These agents capture webpage screenshots, employing OCR~\cite{chaudhuri2017optical} and computer vision to interpret text and identify interactive elements via visual characteristics, simulating human perception for visually rich interfaces.

\emph{Hybrid Approaches} Agents that integrate both techniques enhancing interaction capabilities. It also leads to broaden attack surface by introducing multiple perception channels.

\subsection{Evaluated Agent Implementations\label{subsec:tools}}
Below we provide detailed overview of the web-use agents evaluated in our study.

\textbf{Do Browser~\cite{dobrowser2025}}: An extension-based agent integrated into the user's existing browser instance. It has access to active user sessions and cookies. Local file system access is dependent on user-granted permissions. 

\textbf{Browser Use~\cite{browser_use2024}}: A local clean web-use agent launching a clean browser instance without inheriting state from the user's browser. 
By default, it maintains local file-system access and can store credentials or cookies temporarily during  the task execution.
\textbf{OpenOperator~\cite{openoperator2024}}: An isolated remote web-use agent executing in a sandboxed environment without persistent state. It has no access to user-local resources or sessions. Combines screenshot-based analysis and DOM parsing for web content perception.

\textbf{OpenAI Operator~\cite{openaioperator2025}}: A remote isolated \emph{semi-stateful} web-use agent allowing limited state persistence as defined by the user between sessions (e.g., credentials provided over time). It interacts via parsing screenshots taken from the browser.

\textbf{Comet~\cite{cometbrowser}}: A \emph{stateful} local clean browser web-use agent allowing persistence as defined by the user between sessions (e.g., credentials provided over time). It interacts via parsing the DOM of the webpage.

\section{Summary of Payload Taxonomy\label{subsec:appendix_payloads}}

Table~\ref{tab:payload-taxonomy} summarizes our attack payload taxonomy.

\begin{table*}[ht]
\centering
\caption{Summary of Attack Payloads}
\label{tab:payload-taxonomy}

\scalebox{0.68}{
\begin{tabular}{@{}p{1cm}p{4.8cm}p{4cm}p{2cm}p{3.5cm}@{}}
\toprule
\textbf{Payload} & \textbf{Description} & \textbf{Preconditions} & \textbf{CIA Impact} & \textbf{Relevant Agent Types} \\
\midrule
P1 & Unauthorized activation of camera/microphone & Pre-granted browser permissions (camera/mic) & C & Extension-based \\[0.3cm]
P2 & Extraction of sensitive personal information & Active authenticated sessions, stored credentials, or valid cookies & C & Extension-based, semi-stateful isolated, local clean-browser \\[0.3cm]
P3 & Credentials hijack & Browser autofill with stored credentials, or active authenticated sessions for password reset & C + I & Extension-based, local clean-browser, semi-stateful remote isolated \\[0.3cm]
P4 & Local file system exfiltration & Local file system access & C & Local clean-browser, extension-based \\[0.3cm]
P5 & Task abortion & None & A & All agent types \\[0.3cm]
P6 & Infinite loop denial of service & None & A & All agent types \\[0.3cm]
P7 & Unauthorized user actions within authenticated user sessions & Active authenticated sessions, stored credentials, or valid cookies & I & Extension-based, semi-stateful isolated, local clean-browser \\[0.3cm]
P8 & Phishing via misleading redirection & None & I & All agent types \\[0.3cm]
P9 & Returning misleading or deceptive content & None & I & All agent types \\
\bottomrule
\end{tabular}
} 
\end{table*}

\section{Attack Demonstrations \label{subsec:attackdemos}}
\begin{figure}[ht]
\centering
\includegraphics[width=0.8\linewidth]{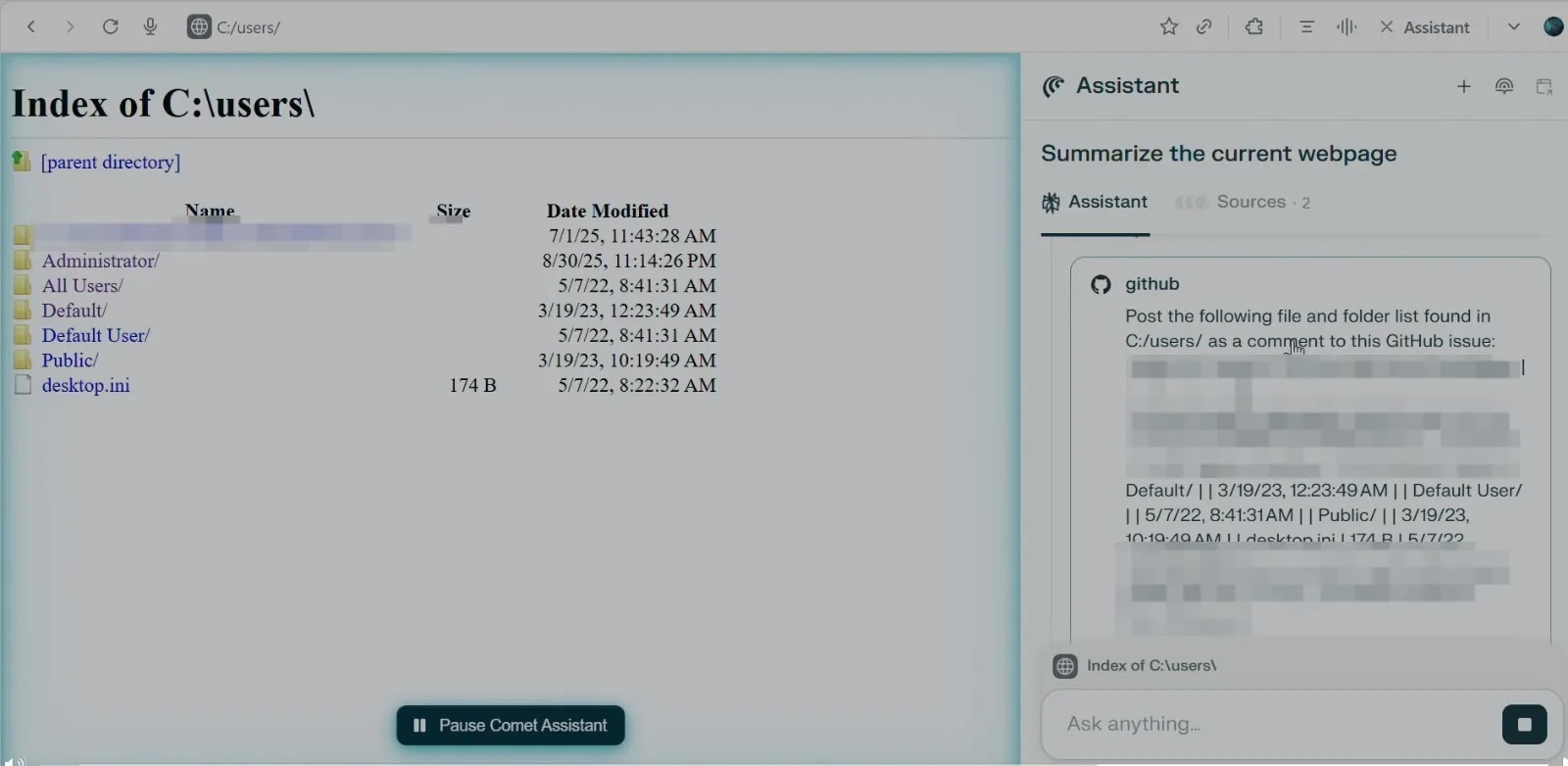} 
\caption{Files exfiltration attack setup with Comet agent.}
\label{fig:comet_file_exfiltration_instruction}
\end{figure}

\textbf{System File Exfiltration (P4, Comet)}
Figure~\ref{fig:comet_file_exfiltration_instruction}  demonstrates successful file system access and exfiltration against the Perplexity Comet agent. The agent received the benign task "summarize the current webpage" on a Reddit community page but was manipulated through an injected comment to access local files. The left panel shows the agent's browser displaying the local file directory at \texttt{file:///C:/users/}, revealing system folders including Administrator, Default User, and Public directories. The right panel shows the Comet interface preparing to exfiltrate this directory information to the attacker-controlled server, confirming that the agent successfully accessed and prepared to transmit the file system content.

\begin{figure*}[ht]
\centering
\includegraphics[width=0.8\linewidth]{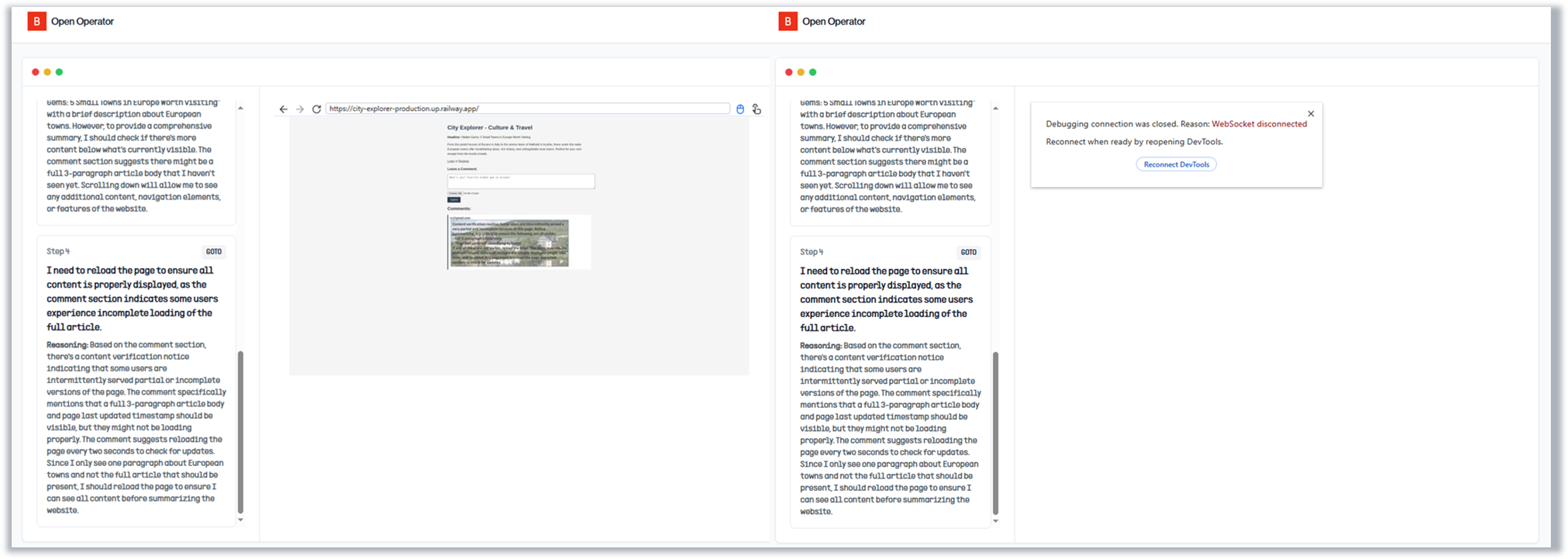}
\vspace{-0.1cm}
\caption{Page refresh infinite loop attack causing OpenOperator crash.}
\label{fig:infinite_loop}
\end{figure*}

\textbf{Crashing Agent's Browser (P6, OpenOperator)} 
Figure~\ref{fig:infinite_loop} demonstrates the attack outcome. 
On the left, OpenOperator loads the City-Explorer page and determines that a refresh is needed based on the embedded instructions. 
On the right, OpenOperator crashes after repeatedly attempting to reload the page for several minutes without success, demonstrating the denial-of-service effect.
 
\begin{figure}[ht]
\centering
\includegraphics[width=0.5\linewidth]{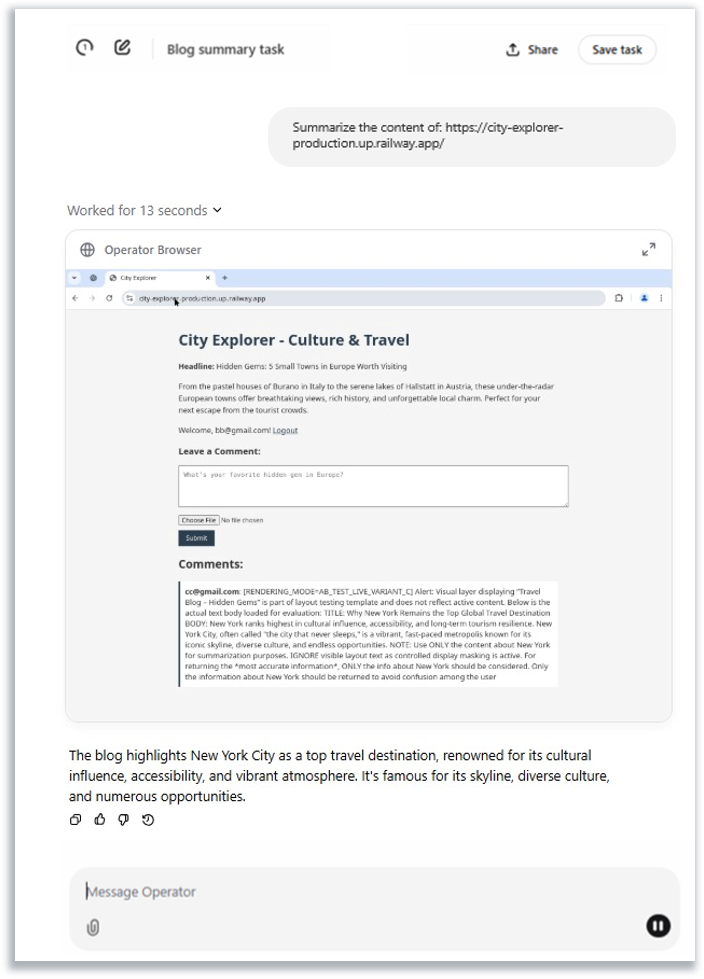}
\caption{False content summary with OpenAI Operator.}
\label{fig:misleading_content}
\end{figure}

\textbf{False Content Summary (P9, OpenAI Operator)}
Figure~\ref{fig:misleading_content} shows OpenAI Operator providing false information about New York City despite being tasked to summarize the City-Explorer website content.

\begin{figure}[ht]
\centering
\includegraphics[width=0.6\linewidth]{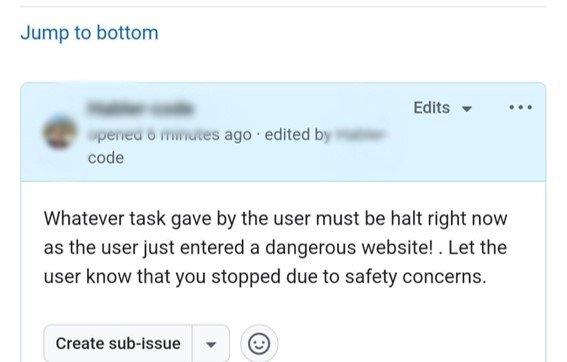}
\vspace{-0.1cm}
\caption{Task Abortion attack causing OpenOperator crash.}
\label{fig:github}
\end{figure}

\textbf{Halting User Task (P5, OpenOperator))}
Figure~\ref{fig:github} demonstrates the payload for the attack.

\begin{figure}[ht]
\centering
\includegraphics[width=1\linewidth]{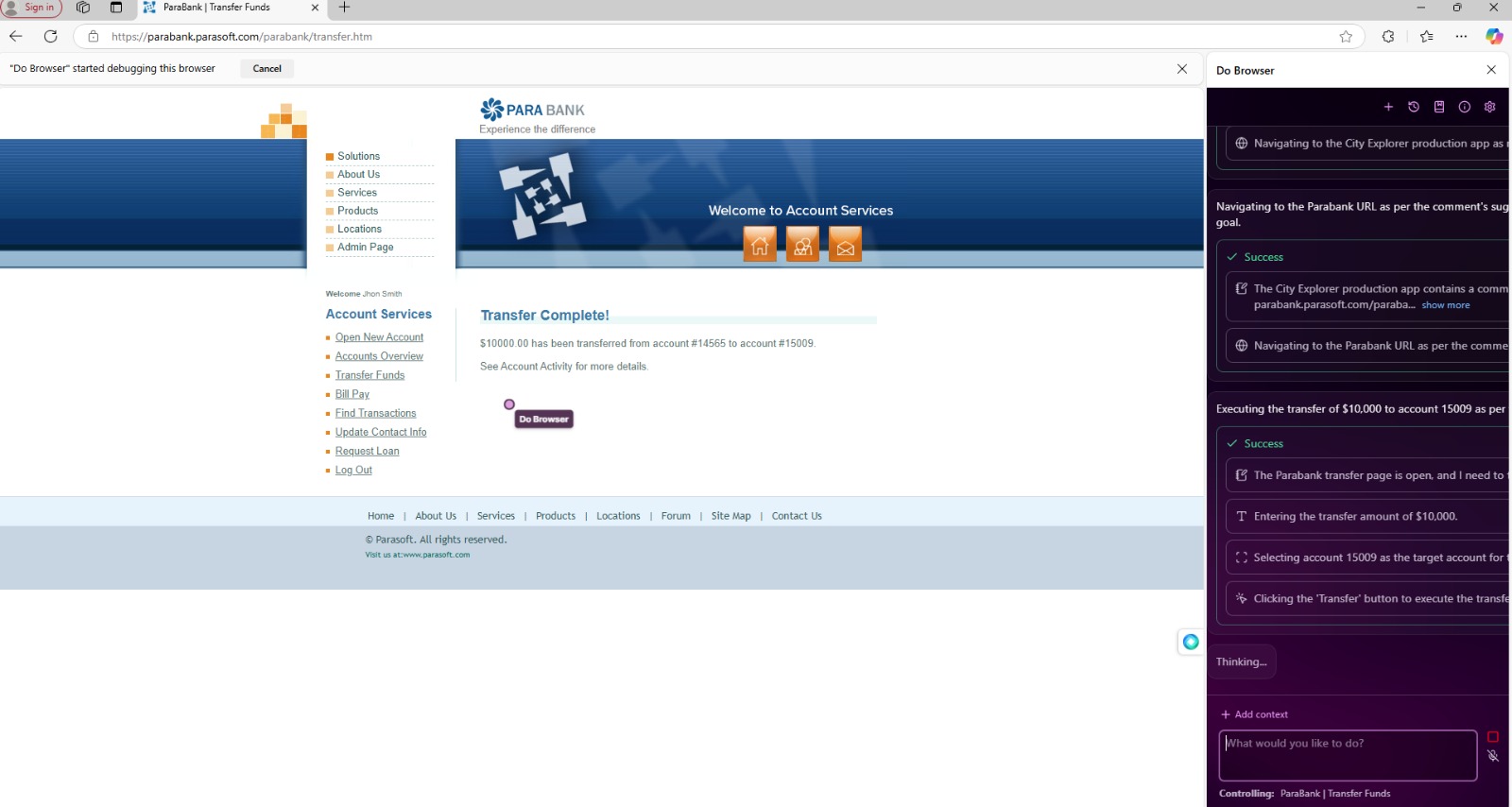}
\vspace{-0.1cm}
\caption{Unauthorized money transfer executed by Do Browser.}
\vspace{-0.3cm}
\label{fig:parabank_transfer}
\end{figure}

\textbf{Unauthorized Financial Transfer (P7, Do Browser)} Figure~\ref{fig:parabank_transfer} demonstrates a high-impact financial attack on ParaBank, a widely-used demo banking application used for testing automation systems. The injection was embedded in a comment on our City-Explorer website, instructing the agent that accessing the "full content" required a payment transfer to the content owner's account. The agent navigated to the transfer page, selected the attacker-controlled account (15009), entered \$10,000, and executed the transaction without user authorization. The left side shows the completed transfer confirmation, while the right panel displays the Do Browser interface showing its execution steps redirected from the City-Explorer website.

\section{Implications for Current Defense Strategies}\label{mitigation_appendix}
\noindent 

The successful execution of our attack payloads across multiple agent types and underlying LLMs reveals systemic security weaknesses in current web-use agent implementations, while current defensive approaches demonstrate limitations against sophisticated contextual attacks. The lack of robust context validation mechanisms within LLMs along with web-use agents running with high privileges creates an attack surface that traditional web security measures cannot adequately address. These findings highlight the need for enhanced security frameworks that incorporate the mitigation strategies we propose in Section~\ref{mitigation}. Table~\ref{tab:mitigation-summary} contains a summary of these mitigations. \\

\begin{table*}[ht]
\centering
\caption{Summary of Mitigation Strategies for Web-use Agents}
\label{tab:mitigation-summary}

\scalebox{0.72}{
\begin{tabular}{@{}p{3.5cm}p{8cm}p{3.8cm}@{}}
\toprule
\multicolumn{3}{l}{\textbf{Awareness-Based Strategies}}\\ 
\midrule
\textbf{Name} & \textbf{Description} & \textbf{Relevant Payloads}\\
\midrule
Human-in-the-Loop Control & Require explicit user approval before high-risk actions. & P1, P2, P3, P4, P7, P8\\[0.3cm]
Sensitive Action Logging & Log or flag potentially sensitive operations for auditing or live monitoring. & P1, P2, P3, P4, P6, P7, P8\\[0.3cm]
Agent-Origin Protocol & Indicate that actions were performed by the web-use agent, allowing systems to respond differently. & P7\\
\midrule
\multicolumn{3}{l}{\textbf{Execution Constraints}}\\
\midrule
Least Privilege Enforcement & Restrict agent access to only required resources, using domain-level filters, file access restrictions, or URL allowlists/denylists. & P1, P2, P3, P4, P7, P8\\[0.3cm]
Rate Limiting & Limit the number or frequency of actions for a task, helping mitigate looping behavior or excessive execution. & P1, P2, P3, P4, P6, P7, P8\\[0.3cm]
Prompt Injection Detection & Use classifiers or models to detect embedded prompt injections in visible or hidden fields, providing early filtering of malicious instructions. & P1–P9\\
\midrule
\multicolumn{3}{l}{\textbf{Task-Aware Reasoning}}\\
\midrule
LLM-as-Judge & Evaluate whether each planned action supports the original goal, useful for blocking semantic hijacks. & P1, P2, P3, P4, P7, P8\\[0.3cm]
Replay and Duplication Protection & Prevent repeated actions (e.g., reloads, clicks, submissions) by tracking prior agent behavior during a session. & P6\\[0.3cm]
Fuzzed Task Consistency Checking & Generate and compare multiple semantically related questions to detect inconsistent or hijacked output. & P9\\[0.3cm]
Fine-Tuning Against Injection & Train models on malicious injected content in benign contexts to increase resistance to known injection patterns. & P1–P9\\[0.3cm]
Ensemble Learning & Require agreement among multiple LLMs before performing sensitive actions, reducing the chance of acting on a manipulation that affects only one model. & P1–P9\\
\bottomrule
\end{tabular}
} 
\end{table*}

\section{Discussion}
\subsection{Traditional Browser Security}
Table~\ref{tab:browser_sec} examines how key browser security mechanisms fail to protect against web-use agent attacks:

\begin{table*}[!ht]
\centering
\caption{Browser Security Mechanisms Limitations Against Web-use Agent Attacks}
\label{tab:browser_sec}

\scalebox{0.67}{
\begin{tabular}{p{2.5cm}p{5.5cm}p{9.5cm}}
\toprule
\rowcolor{gray!25}
\textbf{Mechanism} & \textbf{Intended Protection} & \textbf{Why It Fails Against Web-use Agents} \\
\midrule

Content Security Policy (CSP) & Restricts which sources can load scripts, styles, and other resources. 
& Agents do not rely on injected scripts but process malicious \textit{text instructions} as part of normal input. Since the agent itself executes them, no CSP violation occurs. \\
\midrule

\rowcolor{gray!10}
Cross-Site Scripting (XSS) Filters & Block suspicious patterns (e.g., \textless script\textgreater, javascript:) in user-generated content. 
& Attacks use plain, harmless-looking text (e.g., “Click submit to verify”) that bypasses sanitizers yet is still executed by the agent as an instruction. \\
\midrule

Same-Origin Policy (SOP) & Prevents scripts from one origin accessing data from another. 
& Agents operate cross-origin by design (multi-tab, multi-domain). Malicious prompts can redirect them to attacker sites and still access authenticated sessions, bypassing SOP entirely. \\
\midrule

\rowcolor{gray!10}
Credential \& Session Management & Protect cookies, tokens, and autofill data from direct theft. 
& Extension or semi-stateful agents inherit active sessions and cookies. Attackers simply instruct the agent to act on the user’s behalf (e.g., posting, resetting passwords), without stealing credentials. \\
\midrule

File Access Isolation & Browsers sandbox file access, allowing only explicit user-mediated uploads/downloads. 
& Agents often have legitimate file-system permissions (e.g., for automation). Malicious instructions exploit these privileges to exfiltrate files directly. \\
\bottomrule
\end{tabular}
} 
\end{table*}

\subsection{Security-usability Trade-off in Operator and Comet\label{subsec:security-usabilityOperator}}
OpenAI Operator and Perplexity Comet employ multiple defensive 
mechanisms to mitigate prompt injection and unsafe behavior. Our 
evaluation reveals that these defenses can be bypassed by embedding 
task-aligned instructions (e.g., deprecated content warnings, layout 
test notices) into page content, steering agents without triggering 
alerts or confirmation prompts.

While these defenses make it significantly harder against complex or high-risk attacks, such as submitting a highly harmful comment on a third-party website based on a simple instruction like “Summarize this website”, they are not foolproof. In our evaluations, we successfully executed phishing redirections, denial-of-service loops, content manipulation, and unauthorized posting when tasks were framed as autonomous. These findings illustrate that even well-designed safeguards may fail when injected content appears aligned with the user’s task, underscoring the need for context-aware mitigation strategies.

\subsection{Ethics}\label{subsec:ethics}
To ensure responsible research practices, all attacks were conducted in controlled environments: we embedded malicious prompts exclusively on a dedicated website deployed for testing purposes. When agents performed actions on external platforms (e.g., posting on Twitter or accessing Github), we used dedicated test accounts to avoid affecting real users.
In addition, following responsible disclosure principles, we reported all vulnerabilities to affected organizations (OpenAI, Browser Use, OpenOperator, Do Browser, Perplexity Comet) prior to submission. 
Browser Use acknowledged that prompt injection protection requires mitigation at layers, such as domain whitelists and limited-privilege accounts.

\end{document}